# DAME: a Web Oriented Infrastructure for Scientific Data Mining & Exploration

Massimo Brescia[a,b], Giuseppe Longo[a,b,c], George S. Djorgovski[c], Stefano Cavuoti[b], Raffaele D'Abrusco[d], Ciro Donalek[c], Alessandro Di Guido[b], Michelangelo Fiore[b,] Mauro Garofalo[b], Omar Laurino[e], Ashish Mahabal[c], Francesco Manna[b], Alfonso Nocella[b], Giovanni d'Angelo[b], Maurizio Paolillo[b]

[a]*INAF – Osservatorio Astronomico di Capodimonte, Via Moiariello 16, 80131 Napoli, Italy*

[b]*Dipartimento di Fisica, Università degli Studi Federico II, Via Cintia 26, 80125 Napoli, Italy*

[c] *CALTECH – California Institute of Technology, 1200 East California boulevard, 91125 Pasadena California USA*

[d] *Harvard-Smithsonian Center for Astrophysics, Cambridge USA*

[e]*INAF – Osservatorio Astronomico di Trieste, Via Tiepolo 11, 34143 Trieste, Italy*

**ABSTRACT**

Nowadays, many scientific areas share the same need of being able to deal with massive and distributed datasets and to perform on them complex knowledge extraction tasks. This simple consideration is behind the international efforts to build virtual organizations such as, for instance, the Virtual Observatory (VObs). DAME (DAta Mining & Exploration) is an innovative, general purpose, Web-based, VObs compliant, distributed data mining infrastructure specialized in Massive Data Sets exploration with machine learning methods. Initially fine tuned to deal with astronomical data only, DAME has evolved in a general purpose platform which has found applications also in other domains of human endeavor. We present the products and a short outline of a science case, together with a detailed description of DAMEs main features and architecture.

**KEYWORDS**

Astroinformatics; knowledge discovery in database; data mining; grid; cloud; machine learning; virtual observatory;

## 1. INTRODUCTION

In almost all fields, modern technology allows to capture and store huge quantities of heterogeneous and complex data often consisting of hundreds of features for each record and requiring complex metadata structures to be understood. This has lead to a situation properly described by the famous sentence *"... while data doubles every year, useful information seems to be decreasing, creating a growing gap between the generation of data and our understanding of it"* [1]. A need has therefore emerged for a new generation of software tools, largely automatic, scalable and highly reliable.

Strictly speaking, Knowledge Discovery in Databases (KDD) is about algorithms for inferring knowledge from data and ways of validating the obtained results, as well as about running them on infrastructures capable to match the computational demands. In practice, whenever there is too much data or, more generally, a representation in more than 5 dimensions [2], there are basically three ways to make learning feasible. The first one is trivial: applying the training scheme to a decimated dataset. The second method relies on parallelization techniques, the idea being to split the problem into smaller parts, then solve each using a separate CPU and finally combine the results together [3]. Sometimes this is feasible due to the intrinsic natural essence of the learning rule (such as genetic algorithms; [4]). However, even after parallelization, the algorithm's asymptotic time complexity cannot be improved. The third and more challenging way to enable a learning paradigm to deal with Massive Data Sets (MDS) is to develop new algorithms of lower computational complexity but in many cases this is simply not feasible. DAME copes with this problem by implementing the KDD models under the S.Co.P.E. grid [13] and by taking into account the fact that background knowledge can make it possible to reduce the amount of data that needs to be processed by adopting a learning rule based on the fact that in many cases most of the attributes turn out to be irrelevant when background knowledge is taken into account [3].

Main feature of DAME, however, is its usability which addresses the well known fact that Knowledge Discovery in Databases is a complex process. In most cases, in fact, the optimal results can be found only on a trial and error base by comparing the outputs of different methods. This implies that, for a specific problem, Data Mining requires a lengthy fine tuning phase which is often not easily justifiable to the eyes of a non experienced user.

Such complexity is one of the main explanations for the gap still existing between the new methodologies and the huge community of potential users which fail to adopt them. In order to be effective, in fact, Knowledge Discovery in Databases requires a good understanding of the mathematics underlying the methods, of the computing infrastructures and of the complex workflows which need to be implemented, and most users (even in the scientific community) are usually not willing to make the effort to understand the process and prefer to recur to traditional approaches which are far less powerful but much more user friendly [5].

DAME, by making use of the Web application paradigm, of extensive and user friendly documentation and of a sample of documented and realistic use cases, represents a first attempt to bring the KDD models to the user hiding most of their complexity behind a well designed infrastructure. In this paper we describe the prototype (alpha release) version of the DAME[1] (Data Mining and Exploration) web application, nowadays available under final commissioning which addresses many of the above issues and aims at providing the scientific community with a user friendly and powerful data mining platform.

The paper is structured as follows: in Sections 2 and 3 we outline the main features of DAME and in section 4 we focus on its internal architecture. In section 5 we discuss a scientific use case, presented as an example of results obtained in the astrophysical domain and, finally, in Section 6 we outline the future developments and draw our conclusions.

## 2. GENERAL DESCRIPTION

DAME was initially conceived to work on astrophysical Massive Data Sets data as a tool offered to the community and aimed at solving some of the above quoted problems by offering a completely transparent architecture, a user friendly interface and the possibility to access a distributed computing infrastructure.

DAME starts from a taxonomy of data mining methods (hereinafter called functionalities) and collects a set of machine learning algorithms (hereinafter called models) that can be associated to one or more functionalities depending of the specific problem domain. This association "functionality-model" represents what in the following we shall refer to as "experiment domain".

At a lower level, any experiment launched on the DAME framework, externally configurable through dynamical interactive web pages, is treated in a standard way, making completely transparent to the user the specific computing infrastructure used and the specific data format given as input.

Dimensional reduction, classification, regression, prediction, clustering, filtering, are examples of functionalities belonging to the data mining conceptual domain, in which the various methods (models and algorithms) can be applied to explore data under a particular aspect, connected to the associated functionality scope. In its first implementation the infrastructure prototype has been focused on the classification, regression and clustering functionalities.

A special care was devoted in each single phase of the design and development to produce a complete and exhaustive documentation both technical and user oriented.

---

[1] http://voneural.na.infn.it/

## 3. DESIGN ISSUES

The concept of "distributed archives" is familiar to most scientists. In the astronomical domain, the leap forward was the possibility to organize through the Virtual Observatory (hereafter VObs) [6] the data repositories to allow efficient, transparent and uniform access. In other words, the VObs was intended to be a paradigm to use multiple archives of astronomical data in an interoperating, integrated and logically centralized way, so to be able to "observe" and analyze a virtual sky by position, wavelength and time [38]. In spite of some underscoping, the VObs can still be considered as an extension of the classical computational grid, in the sense that it fits the data grid concept, being based on storage and processing systems, together with metadata and communications management services. The link between data mining applications and the VObs data repositories is currently still under discussion since it requires (among the other things) the harmonization of many recent achievements in the fields of VObs, grid, cloud, HPC (High Performance Computing), and Knowledge Discovery in Databases.

DAME was conceived to provide the VObs with an extensible, integrated environment for data mining and exploration. In order to do so, DAME had to:

- Support the VObs standards and formats, especially for data interoperability;
- To abstract the application deployment and execution, so to provide the VObs with a general purpose computing platform taking advantage of the modern technologies.

In order to gradually accomplish such requirements, DAME intended to give the possibility to remotely interact with data archives and data mining applications via a simple web browser [7]. Thus, with web applications, a remote user does not require to install program clients on his desktop, having the possibility to collect, retrieve, visualize and organize data, configure and execute mining applications through the web browser and in an asynchronous way. An added value of such approach being the fact that the user does not need to directly access large computing and storage power facilities. He can transparently make his science by exploiting computing networks and archives located worldwide, requiring only a local laptop (or a personal smartphone) with a network connection.

For what the DAME functionalities are concerned, we need to distinguish between supervised and unsupervised algorithms. In the first case, we have a set of data points or observations, for which we know the desired output, class, target variable, or outcome. The outcome may take one of many values called "classes" or "labels". The target variable provides some level of supervision in that it is used by the learning algorithm to adjust parameters or make decisions that allow it to predict labels for new data. Finally, when the algorithm is predicting labels to observations we call it a classifier. Some classifiers are also capable of providing a probability for a data point to belong to a given class and are often referred to as probabilistic model or regression, not to be confused with statistical regression model [8].

In the unsupervised case, instead of trying to predict a set of known classes, we are trying to identify the patterns inherent in the data, without outcomes or labels. In other words, unsupervised methods try to create clusters of data that are inherently similar.

Strictly connected with the dichotomy between supervised and unsupervised learning, in the DAME data mining infrastructure, the choice of any machine learning model is always accompanied by the functionality domain. To be more precise, several machine learning models can be used in the same functionality domain, since it represents the experiment domain context in which it is performed the exploration of data.

In what follows we shall therefore adopt the following terminology:

- **Data mining model**: any of the data mining models integrated in the DAME suite. It can be either a supervised machine learning algorithm or an unsupervised one, depending on the available Base of Knowledge (BoK, i.e. the set of training or template cases available) and the scientific target of the user experiment;
- **Functionality**: one of the functional domains in which the user wants to explore the available data (for example, regression, classification or clustering). The choice of the functionality target can limit the choice of the data mining model;
- **Experiment**: it is the scientific pipeline (including optional pre-processing or preparation of data) and includes the choice of a combination of data mining model and a functionality;

- **Use Case**: for each data mining model, different running cases are exposed to the user . These can be executed singularly or in a prefixed sequence. Being the models derived from the machine learning paradigm [9], each has training, test, validation and run use cases, in order to, respectively, perform learning, verification, validation and execution phases. In most models there is also the "full" use case, that automatically executes all listed cases as a sequence.

| MODEL | CATEGORY | FUNCTIONALITY |
|---|---|---|
| MLP (Multi Layer Perceptron) with Back Propagation learning rule | Supervised | Classification, Regression |
| MLP with Genetic Algorithms learning rule | Supervised | Classification, Regression |
| MLP with Quasi Newton learning rule | Supervised | Classification, Regression |
| SVM (Support Vector Machine) | Supervised | Classification, Regression |
| MLC (Multi Layer Clustering) | Unsupervised | multiple hierarchical clustering |
| SOM (Self Organizing Maps) | Unsupervised | Clustering, image segmentation |
| PPS (Principal Probabilistic Surfaces) | Unsupervised | Dimensional reduction, |
| DAME-NExTII astronomical pipeline | Unsupervised | Image Segmentation, Clustering |

*Tab. 1 – the data mining models and functionalities made available in DAME*

The functionalities and models already and incoming available in DAME, are listed in Table 1.
Depending on the specific experiment and on the execution environment, the use of any of the above models can take place with a more or less advanced degree of parallelization. All the models require some parameters that cannot be defined a priori, thus causing the necessity of iterated experiment interactive sessions in order to find the best tuning. For this reason not all the models can be developed under the MPI (Message Passing Interface) paradigm [10]. But the possibility to execute more jobs at once (specific grid case) intrinsically exploits the multi-processor architecture. As for all data mining models integrated in the DAME framework, there is instanced a specialized plugin (called DMPlugin[2]) for each couple model-functionality, that includes the specific parameter configuration. In this way, it is also possible to easily replace and/or upgrade models and functionalities without any change or modification to the DAME software components [27].

## 4.  THE DAME ARCHITECTURE

The DAME design architecture is implemented following the standard LAR (Layered Application Architecture) strategy, which leads to a software system based on a layered logical structure, where different layers communicate with each other via simple and well-defined rules:

- Data Access Layer (DAL): the persistent data management layer, responsible of the data archiving system, including consistency and reliability maintenance.
- Business Logic Layer (BLL): the core of the system, responsible of the management of all services and applications implemented in the infrastructure, including information flow control and supervision.
- User Interface (UI): responsible of the interaction mechanisms between the BLL and the users, including data and command I/O and views rendering.

A direct implication of the LAR strategy adopted in DAME is the Rich Internet Application (RIA) [11], consisting in applications having traditional interaction and interface features of computer programs but usable via simple web browsers, i.e. not needing any installation on user local desktop. RIAs are particularly efficient in terms of interaction and execution speed.
By keeping this in mind, the main concepts behind the distributed data mining applications implemented in the DAME Suite are based on three issues:

- Virtual organization of data: this is the extension of the already remarked basic feature of the VObs;

---
[2] http://voneural.na.infn.it/dmplugin.html

- Hardware resource-oriented: this is obtained by using computing infrastructures, like grid, which enable parallel processing of tasks, using idle capacity. The paradigm in this case is to obtain large numbers of instances running for short periods of time;
- Software service-oriented: this is the base of typical cloud computing paradigm [12]. The data mining applications implemented runs on top of virtual machines seen at the user level as services (specifically web services), standardized in terms of data management and working flow.

The complete Hardware infrastructure of the DAME Program is shown in Fig. 1, where the grid sub-architecture provided by the S.Co.P.E. supercomputing facility [13], [27] is incorporated into the more general cloud scheme, including a network of multi-processor and multi-HDD PCs and workstations, each of them dedicated to a specific function.

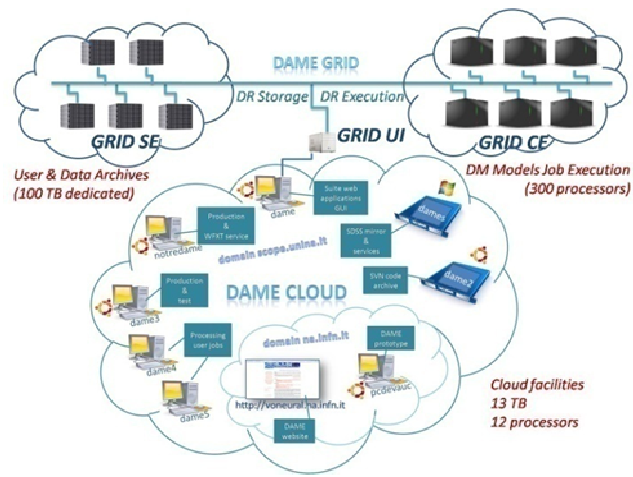

**Fig.1 – The DAME hybrid computing architecture**

There are two sub-networks addressable from an unique access point, the website, which provides an embedded access to the user to all DAME web applications and services. The integrity of the system, including the grid public access, is guaranteed by a registration procedure, giving the possibility to access all facilities from just one account. In particular, a robot certificate is automatically handled by the DAME system to provide transparent access to the S.Co.P.E. grid resources [14].

Depending on the computing and storage power, requested by the job and by the processing load currently running on the network, an internal mechanism redirects the jobs to a job-queue in a pre-emptive scheduling scheme. The interaction with the infrastructure is completely asynchronous and a specialized software component (DR, DRiver) has the responsibility to store off-line job results in the user storage workspaces, that can be retrieved and downloaded in subsequent accesses.

This hybrid architecture, renders it possible to execute simultaneous experiments that gathered all together, bring the best results. Even if the individual job is not parallelized, we obtain a running time improvement by reaching the limit value of the Amdahl's law (N):

$$\frac{1}{(1-P) + \frac{P}{N}}$$

where P is the fraction of a program that can be made parallel (i.e. which can benefit from parallelization), and (1 − P) is the fraction that cannot be parallelized (remains serial), then the resulting maximum speed-up that can be achieved by using N processors is obtained by the law expressed above.

For instance, in the case of the AGN (Active Galactic Nucleus) classification experiment detailed in [15], [26], each of the 110 jobs runs for about a week on a single processor. By exploiting the grid, the experiment running time can be reduced to about one week instead of more than 2 years.

From the software point of view, the baselines behind the engineering design of DAME Suite were:

- *Modularity*: software components with standard interfacing, easy to be replaced;
- *Standardization*: basically, in terms of information I/O between user and infrastructure as well as between software components (in this case based on XML-schema);
- *Hardware virtualization*: i. e. independent from the hardware deployment platform (single or multi processor, grid etc.);
- *Interoperability*: by following VO requirements;
- *Expandability*: many parts of the infrastructure require to be increased along its lifetime. This is particularly true concerning computing architecture, framework capabilities, GUI (Graphical User Interface) features, data mining functionalities and models (this also includes the integration within the system of end user proprietary algorithms);
- *Asynchronous interaction*: the end user and the client-server mechanisms do not require a synchronous interaction. In other words, the user is not constrained to remain connected after launching an experiment in order to wait for the end of execution. Moreover, by using the Ajax (Asynchronous Javascript and XML [19]) mechanism, the web applications can retrieve data from the server asynchronously in background without interfering with the display and behavior of the existing page.
- *Language-independent Programming*: this basically concerns the API (Application Programming Interface) forming the data mining model libraries and packages. Although most of the available models and algorithms were internally implemented, this is not considered as mandatory (it is possible to re-use existing tools and libraries, integration of end user tools etc.). So far, the Suite provided a Java based standard wrapping system to achieve the standard interface with multi-language APIs;
- *Distributed computing*: the components can be deployed on the same machine as well as on different networked computers;

The DAME software architecture is based on five main components: Front End (FE), Framework (FW), Registry and Data Base (REDB), Driver (DR) and Data Mining Models (DMM). The scheme in Fig. 2 shows the component diagram of the entire suite.

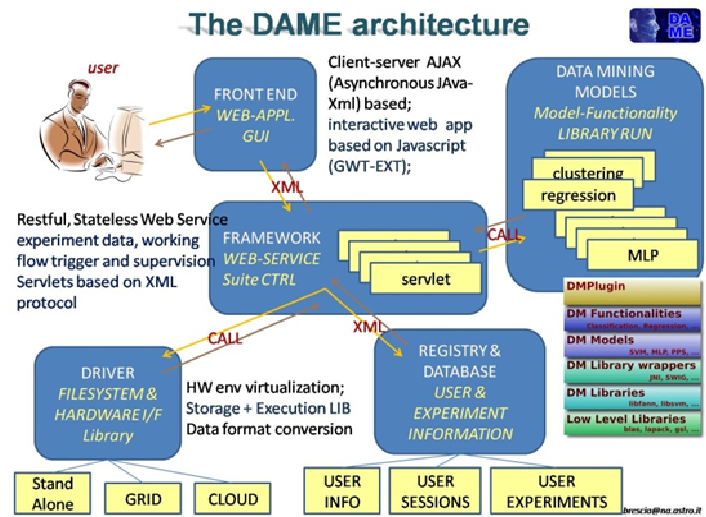

**Fig.2 – The DAME software architecture**

The Front End (FE) is the component directly interfacing the end user with the infrastructure (through the web browser). and it can be considered structured in two main parts (also addressable as a "client" and a "server", even though both of them are resident on the remote side with respect of the user): the main GUI (Graphical User Interface) of the Suite and the internal interface (hereinafter Front End Server) with the inner infrastructure.

The GUI part is based on dynamical WEB pages, rendered by the Google Web Toolkit (GWT, [16]), and it interfaces the end users with the applications, models and facilities needed to launch scientific experiments (Fig. 3). The GWT technology approach was chosen in order to write web applications by using platform-independent Java language, then converting them, at the end of the embedding process, into Javascript code, compliant with HTML browser code.

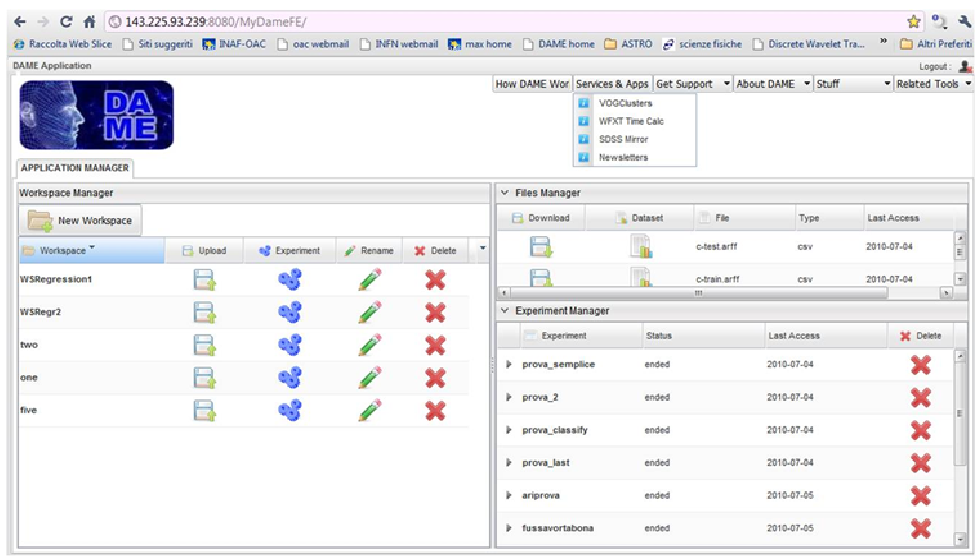

**Fig.3 – The DAME Graphical User Interface (α release)**

This choice was dictated by the following considerations:

- The use of Java code makes it easier to be compliant with OOP (Object Oriented Paradigm). Moreover, Java is a platform-independent code which allows a high level of portability and easy maintenance;
- The embedded conversion from Java to Javascript and HTML code makes the web application automatically compliant with all commonly available web browsers. Moreover, it permits to reduce the developer learning curve, because it basically requires the Java language knowledge only. The GWT library implemented in DAME is based on SmartGWT, internally based on a Javascript SmartClient library [17]. Here the Javascript code, self-generated by the GWT compiler, makes calls to the SmartClient library that are rendered in the browser in the easiest way. The GWT has the advantage of being a young technology, which implies that it will likely be supported by their developers for many years (it is currently adopted for all Google services and resources, like Gmail, Google Map etc.); Moreover GWT offers a huge multiplicity of user interface components and supports flexibility in the custom design of GUI based on specific requirements;
- The communication client-server is based on the RPC (Remote Procedure Call) mechanism [18], that gives the opportunity to shift all user interface business logic on client side. This implies improved performances, as well as the reduction of transmission bandwidth on the web server and of the overload on server side. At the end it achieves a better (in terms of quick response) behavior of the communication with the end user. In DAME, with the exception of the file download and upload tasks, which require a specific procedure, all other client-server communication tasks have been implemented with this mechanism;

The Front End Server is the "embedded" part of the component, directly interacting with the inner side of the infrastructure. One of the technologies employed in order to be compliant with the mentioned Rich Internet Application (RIA) requirement is Ajax [19], that permits an asynchronous use of the Javascript technique through an interface based on XML. It basically optimizes the information flow between a client and a server, in terms of efficiency and speed. One of the advantages of a Front End component based on the Ajax technology is that for all user commands and actions, the processing of the request (a simple collection or manipulation of data) is done on the server side (Fig. 4).

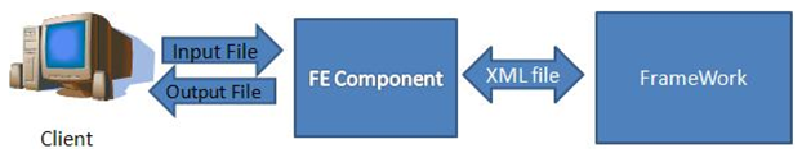

**Fig.4 – The Front End as business logic module between user and inner infrastructure**

The Server is able to provide a reply to the client, as a standard HTML page, in an asynchronous way, i.e. without waiting for the process completion, by avoiding the usual delays of standard web applications.
The FE also provides the interactive /asynchronous check of experiment status, the download of final or partial results and the data mining model-dependent output graphics and charts eventually selected by the user.
The main tasks performed by the Front End Server are:

- To check the status of uploaded files;
- To check the correctness of the parameters inserted by the user for a given experiment;
- To send to the Framework the processing request of an experiment together with information about the user and the working session;
- To check the status of the output files and to flow them to the GUI when they are ready for download (also partial output data are foreseen in the case of interactive sessions);

The safety policy is based on the authentication procedure and on the assignment of a SSID (Security Session ID) to each login session which is cross-checked at each operation request. When user jobs are scheduled to be executed on the grid, the safety is guaranteed by the certification mechanisms intrinsically embedded in the infrastructure and completely transparent to the end user.
The Framework (FW) is the core of the Suite. It handles all communication flow from/to FE (i.e. the end user) and the other components, in order to register the user, to show the working session information, to configure and execute all the user experiments, to report output and status/log information about the applications which are either running or already finished. Main tasks performed and directly managed by the FW component inside the DAME infrastructure are based on XML type document compiling and exchange with other components. These tasks are:

- Delivering data related to the data mining functionalities to the FE, that is the description list and the single description documents;
- Authenticating users on behalf of the FE, by working together with the Database Management System (DBMS);
- Registering users on behalf of the FE, by working together with the DBMS;
- Delivering a working session list, a file list and the meta-data description to the FE;
- Receiving files from the FE and storing them on the FileStore and the DataBase by working with the Driver component (DR) and the DBMS;
- Uploading files sent by the FE on the FileStore, by working with the DR;
- Monitoring the status of components and experiments;
- Providing the FW administrator management tools to install, enable and disable functionalities, as well as other administration tasks that may be useful;
- Loading and configuring a plugin, after receiving a configuration file related to a functionality from the FE. After that the FW has to communicate with the DR start the job execution, and give a feedback to the Plugin every time its status is updated;

The FW component is based on a restful web service [20], that is a client-server architecture in which web services are viewed as resources and are identified by their URL (Uniform Resource Locator) address. A client that wants to gain access to a resource can use common http methods like "get" and "post" on the URL that identifies the desired resource.

A restful environment is completely stateless, that is every interaction is atomic. We decided to use Java Servlets to implement the RESTful Web Service, which allows to add dynamic contents to a Web Server using the Java platform. A Servlet is an object which accept a request from a client and creates a response based on that request [21]. The advantages of using servlets are their fast performance and ease of use combined with more power over traditional Common Gateway Interface (CGI), in terms of server overload and database access [22]. The servlet-based communication between FE and FW is managed by XML files, through the JDOM (Java Document Object Model) package, an open source library, designed to create and develop an XML document in simple and direct way [23]. An example of this mechanism is shown in Fig. 5.

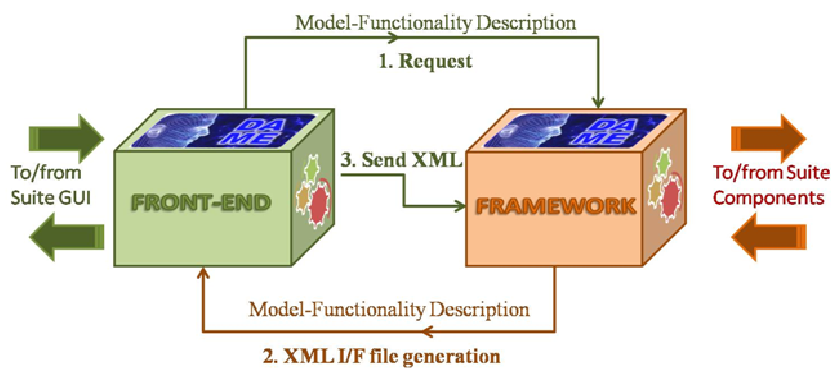

**Fig.5 – Example of servlet interface between Front End and Framework**

The component REgistry & DataBase (REDB) is the base of knowledge repository for the Suite. It is a registry in the sense that it contains all information related to user registration and accounts, as well as to his working sessions and experiments. The REDB needs to manage the user registration, authentication, working sessions, experiments and files. The DBMS (DataBase Management System) part manages data and relationships about users, sessions, experiments, functionalities and files. The component architecture is based on the following entities: a relational DBMS composed by a DB Server and a text interface client that supports the SQL (Structured Query Language) syntax; a JDBC (Java DataBase Connectivity) Driver and a JDBC API (Application Programming Interface) and, finally, the user interface (Fig. 6).

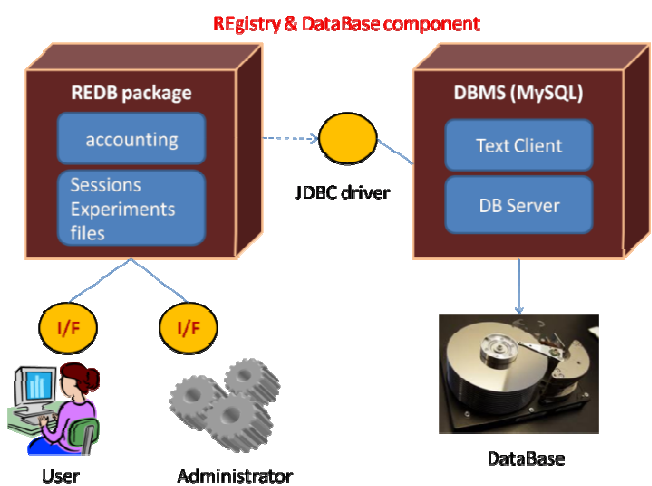

**Fig.6 – The REgistry & DataBase component architecture**

The REDB component is also a Database containing information about experiment input/output data and all temporary/final data coming from user jobs and applications.

The component called DRiver (DR) is the package responsible of the physical implementation of the HW resources handled by other components at a virtual level (Fig. 7).

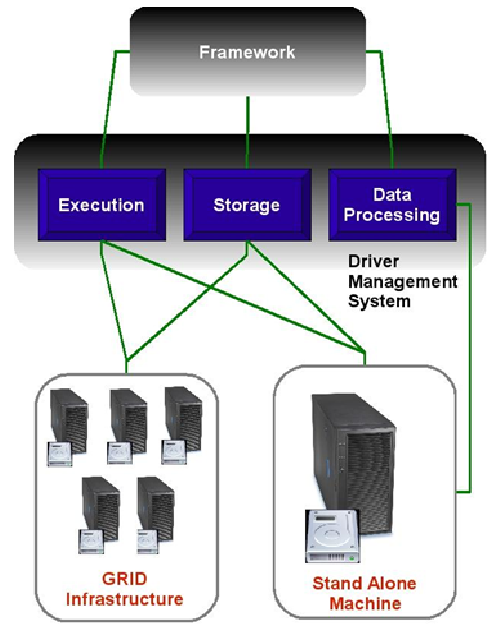

**Fig. 7 – The DRiver component as a virtualized computing environment**

It allows the abstraction of the real platform (hardware environment and related operative system calls) from the rest of the Suite software components, including also I/O interface (file loading/storing), user data intermediate formatting and conversions (ASCII, CSV, FITS, VO-TABLE) through STIL library [24], job scheduler, memory management and process redirection.

At low level, any experiment launched on the DAME framework, externally configurable through dynamical interactive web pages, is treated in a standard way, making completely transparent to the user the specific computing infrastructure used and specific data format given as input.

The final product is a distributed data mining infrastructure, perfectly scalable in terms of data dimension and computing power requirements. In practice, a specific software module of the Suite, called DRiver Management System (DRMS), that is a sub-system of the DRIVER (DR) component has been implemented to delegate at runtime the choice of on which computing infrastructure should be launched the experiment. Currently, the choice is between grid or stand alone multi-thread platform, that could be replaced also by a cloud infrastructure,

The mechanism is simple, being based on a threshold-based evaluation of the input dataset dimensions and the status of grid job scheduler at execution startup time. This reduces both the execution time on a single experiment and the entire job execution scheduling.

The component DMM is the package implementing all data processing models and algorithms available in the Suite. They are referred to supervised/unsupervised models, coming from Soft Computing, Self-adaptive, Statistical and Deterministic computing algorithms [25]. It is structured by means of packages or libraries (API) referred to the following items:

- Data mining models libraries (as listed in Tab. 1);
- Visualization tools;
- Statistical tools;
- List of exposed functionalities (as listed in Tab.1);

- Custom libraries required by the user;

The basic requirements and related solutions of the DMM component are:

- Implementation strategy oriented to the functionalities, i.e. classes of functionalities;
- The possibility of use functionalities with more than one model, thus avoiding code duplication, i.e. by using a specific design pattern (the known *bridge* pattern in this case);
- Separation between Supervised and Unsupervised models, i.e. by means of a specific interface hierarchy of classes;
- A common interface for all the models. By this way it is possible to implement the same interface for all the models and algorithms;

So far, the DMM is composed by several models divided by category. Currently, there are two categories, "Supervised" and "Unsupervised" algorithms divided into three functionalities: classification, regression and clustering. The DMM design strategy is modular, i.e. easy to be extended by adding new functionality categories and/or new data mining models and algorithms, without requiring structural modifications to the high level class hierarchy [27].

DMM consists internally of an interface, called *DMMInterface*, that represent a generic data mining model that can be added to the suite. The Bridge design pattern has been implemented to save code and to achieve the optimized and coherent combination of classes belonging to functionality and model categories, respectively, (Fig. 8).

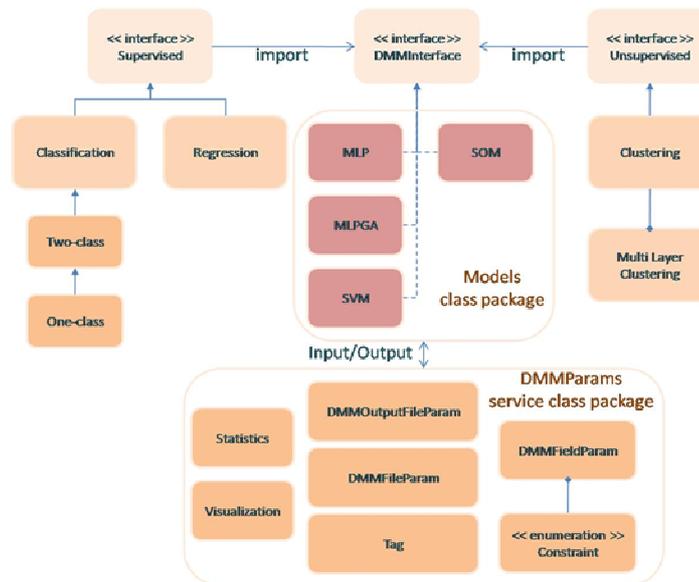

**Fig. 8 – The DMM Class Diagram implementing the Bridge design pattern**

All the data mining models and related functionalities, made available inside the DAME suite, have been based on their own pre-integrated plugins. The main advantage is the possibility to expand the DAME suite by adding new functionalities and data mining models, without affecting the internal structure of the program [27].

The DMPlugin allows a generic user to integrate its own data mining application/routine without the need to understand the programming rules and standard of the DAME suite. The advantage being the possibility to exploit both computing power and storage infrastructures as well as the other facilities offered by the suite.

# 5. AN APPLICATION OF DAME

Since DAME relies upon well known algorithms, we shall not enter into the details of the testing and validation of the individual methods but, in order to exemplify the capabilities of the DAME infrastructure, we shall focus on the results of a realistic scientific experiment.

As it was already mentioned, DAME has been initially conceived for astrophysical applications and has been extensively used for a variety of science cases, such as the evaluation of photometric redshifts of galaxies [28], QSO's (Quasi Stellar Object) [29] and the classification of Active Galactic Nuclei [15], [26].

| Model | Multi Layer Perceptron | | |
|---|---|---|---|
| Topology | Feed forward (three layers) | | |
| Input nodes | Max | 11 | Complete dataset |
| | Min | 4 | Pruning on optical features |
| | Optical features only | 7 | For experiments with optical features only |
| Hidden nodes | Depending on number of features | | |
| | Max | 23 | |
| | Min | 15 | |
| Output nodes | Based on crispy classification | | |
| | Number in BoK | 1 | (0 if no GC, 1=else) |
| | Number in model | 2 | Automatically generated for internal statistics reasons |
| Activation Functions | Input layer | Linear | |
| | Hidden | Not linear, tanh (hyperbolic tangent) | |
| | Output | SOFTMAX | |
| **Learning rule information** | | | |
| Output error type | Cross entropy | | |
| Training mode | Batch | | |
| Training algorithm | Quasi Newton | | |
| QNA Implementation rule | Based on L-BCFG method (L stands for limited memory) | | |
| QNA Parameters | Value | Meaning | |
| Decay | 0,001 | Weight decay during gradient approximation | |
| Restarts | 20 | Random restarts for each approx. step | |
| Wstep | 0,01 | Stopping threshold (min error for each step) | |
| Maxlts | 1500 | Max number of iterations for each approx. step | |

*Tab. 2 – Summary of the experiment*

In this section we shortly outline the results of another experiment performed with the α-release of the platform and concerning the identification of globular clusters in external galaxies. Further details will be found in [31], [32]. For the benefit of non astronomer readers, we shall just point out that Globular Clusters (GCs) are almost spherical, massive stellar systems orbiting in the external halo of galaxies. The study of the GCs populations in external galaxies requires the use of wide-field, multi-band photometry and, in order to minimize contamination from fore/background objects and to measure some of the GC properties (size, core radius, concentration, binary formation rates) high angular resolution data are required [30].

The detection of GCs relies basically on two aspects: the shape of the image (which differs from the instrumental Point Spread Function or PSF) and the colors (i.e. the ratio of observed fluxes at different wavelengths). The shape allows to disentangle large systems from stars (which are PSF-like), while the colors are needed to disentangle GCs from other extended systems such as background galaxies (Fig. 9).

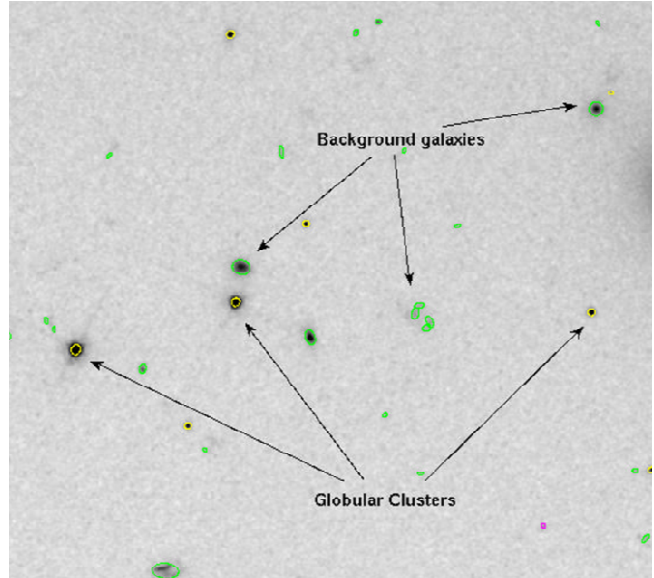

**Fig. 9 – Section of Hubble ACS image used to detect Globular Clusters around N1399. GCs (in yellow) are difficult to distinguish from background galaxies (in green) based only on single band images.**

The supervised learning experiment presented in what follows, regarded the attempt to identify GCs in single band wide field images obtained with the Hubble Space Telescope for the galaxy NGC1399, using the base of knowledge ("true" GCs) provided in [31], [32]. The advantage being that single band data are much less expensive in terms of observing time, and thus easier to obtain than multi-band ones. The input (see Table 3) features were of two types: optical (measured fluxes and moment of the light distribution) and structural (derived from a King model fit, commonly used to describe GC profiles). Optical parameters were measured for 12915 objects from a single band deep image of the galaxy NGC1399, while structural parameters were measured for a subsample of 4590 sources [31], [32].

The Book of Knowledge (BoK) used to train the model was obtained by using multi-wavelength information (color selection). The total amount of objects in the BoK was 2100 objects, having both optical color and structural information: 1219 "true" GCs and 881 "false" GCs.

The machine learning supervised model which obtained the best recognition performances was the Multi Layer Perceptron (MLP) trained by the Quasi Newton Approximation (QNA) learning rule [33], [34], implemented with the known L-Broyden–Fletcher–Goldfarb–Shanno (L-BCFG) [35], where L stands for "Limited memory" version of the algorithm, available in the incoming DAME β release, configured in three layers (input-hidden-output). With this method we performed the series of experiments summarized in Table 2. Using all features the best result led to a performance of 98.33%. It needs to be stressed, however, that a feature significance analysis performed by rejecting one feature at the time (pruning), showed that the exclusion of feature 11 does not significantly degrade the performances (97.95%), [36].

| *Id* | *LABEL* | *Type* |
|---|---|---|
| *1* | *MAG_ISO* | *Optical* |
| *2* | *MAG_APER1* | *Optical* |
| *3* | *MAG_APER2* | *Optical* |
| *4* | *MAG_APER3* | *Optical* |
| *5* | *KRON_RADIUS* | *Optical* |
| *6* | *ELLIPTICITY* | *Optical* |
| *7* | *FWHM* | *Optical* |
| *8* | *CENTRAL SURFACE BRIGHTNESS* | *Structural* |
| *9* | *CORE_RADIUS* | *Structural* |
| *10* | *EFFECTIVE RADIUS* | *Structural* |
| *11* | *TIDAL RADIUS* | *Structural* |
| *12* | *GC CLASSIFICATION* | ***Target*** |

*Tab. 3 – Composition of dataset features for experiments*

More in detail, concerning the best performance case (the dataset with 2100 samples, including both optical and structural features), the reported performance of 98.33% is hence referred to the following model output:
- 1203 TRUE GCs correctly identified;
- 862 FALSE GCs correctly identified;

The results can also be expressed also in terms of completeness and purity of the experiment:

- 1203 TRUE GCs recognized out of 1219 samples imply a completeness of 98.69%;
- 19 FALSE GCs were wrongly considered as TRUE, so far contaminating the output dataset. It hence results with a purity of 98.44% (1.56% of contamination).

Finally, as previously mentioned, a complete pruning phase was performed on all 11 features of the BoK (7 optical plus 4 structural features), obtaining a relevance percentage (in terms of correlation contribute in each sample) for all features. The fact that feature 11 (TIDAL RADIUS) carries almost no relevant information can be easily understood on the basis that the tidal radius of globular clusters is not well determined, thus resulting in a very low contribution in terms of correlation information [36].

## 6. CONCLUSIONS AND FUTURE DEVELOPMENTS

DAME is an evolving platform and new modules and specific workflows as well as additional features are continuously added. The modular architecture of DAME can also be exploited to build applications, finely tuned to specific needs. Examples available so far and accessible through the DAME website, being VOGCLUSTERS (Virtual Observatory Globular Clusters), a VObs web application aimed at collecting and make available all existing data on galactic globular clusters for data and text mining purposes, and NExt-II (Neural Extractor) for the segmentation of wide field astronomical images.

Although all models reported in Tab. 1 have already been tested on scientific cases or, more simply validated in terms of algorithm debugging, their public availability in the DAME Suite has to follow the internal work package schedule. The first model listed in Tab.1 (MLP with BP rule) is implemented in the α release of the Suite, already available at http://voneural.na.infn.it/alpha_info.html. While the first 6 items, reported in Tab.1, have been made available with the official β release, of November 2010 (http://voneural.na.infn.it/beta_info.html). Finally the complete set of models (and functionalities) is scheduled to be released in the summer of 2011.

Other foreseen developments are:

- The implementation of an interoperable platform to join data analysis and exploration features of DAME and KNIME projects, [37]. This could achieve the possibility to gain mutual enhancements between the two infrastructures, based, respectively, on distributed server-side and desktop application paradigms;
- Introduction of MPI (Message Passing interface) technology in the Framework and DBMS components of DAME, by investigating its deployment on a multi-core platform, based on GPU+CUDA computing technique [39]. This could improve computing efficiency in data mining models, such as Genetic Algorithms, naturally implementable in a parallel way.

In conclusion, we are confident that DAME may represent what is generally considered an important element of e-science: a stronger multi-disciplinary approach based on the mutual interaction and interoperability between different scientific and technological fields (nowadays defined as X-Informatics, such as Astro-Informatics). Such an approach may have significant implications in the Knowledge Discovery in Databases process, where even near-term developments in the computing infrastructure which links data, knowledge and scientists will lead to a transformation of the scientific communication paradigm and will improve the discovery scenario in all sciences.

*Acknowledgements*

All partners acknowledge the financial support of the Italian Ministry of Foreign Affairs for the Italy-USA bi-lateral grant "Building an e-science Data Mining Infrastructure", the European Union through the projects VO-Tech and VO-AIDA and the Ministry of University and Research through the PON S.Co.P.E. GRID project. SGD, AAM, and CD acknowledge a crucial support from the Fishbein Family Foundation, and a partial support from the NASA grant 08-AISR08-0085.